\documentclass[3p,preprint,twocolumn]{elsarticle}

\usepackage{hyperref}
\usepackage{bm}
\usepackage{placeins}
\usepackage{array}
\usepackage{caption}
\usepackage{cancel}
\usepackage{makecell} 
\usepackage{textcomp, gensymb}
\usepackage{amsmath}
\usepackage{booktabs,makecell,multirow,array,colortbl,dcolumn,stfloats} 

\usepackage{framed} 
\usepackage{multicol} 
\usepackage{nomencl} 
\makenomenclature
\renewcommand*\nompreamble{\begin{multicols}{2}}
\renewcommand*\nompostamble{\end{multicols}}

\usepackage{color}
\usepackage[size=footnotesize]{todonotes} 
\usepackage{relsize}
\usepackage[utf8]{inputenc}
\usepackage{mathtools}
\usepackage{amsmath}
\usepackage{amssymb}

\journal{International Journal of Fatigue}

\begin{document}
\sloppy

\begin{frontmatter}

\title{Identification and characterization of three-dimensional crack propagation mechanism in the Aluminium alloy AA2024-T3 using high-resolution Digital Image Correlation}

\author[mymainaddress]{Vanessa Schöne\corref{mycorrespondingauthor}}
\cortext[mycorrespondingauthor]{Corresponding author}
\ead{vanessa.schoene@dlr.de}
\author[mymainaddress]{Florian Paysan}

\author[mymainaddress]{Eric Breitbarth}

\address[mymainaddress]{Institute of Materials Research, German Aerospace Center (DLR), Köln, Germany}

\begin{abstract}
Fatigue crack growth is usually a three-dimensional problem, but it is often simplified to two dimensions to reduce complexity. However, this study investigates the relationships between microscopic effects such as crack kinking, shear lips, and plasticity that are present in reality. Therefore, crack propagation tests were carried out on 2-mm-thick MT-160 specimens of AA2024-T3 sheet material in L-T and T-L orientation. Using high-resolution digital image correlation (DIC), the plastic zone was identified and measured on the samples surface. The fracture surfaces were then digitised and their 3D shape characterised. Finite element simulations confirm the presence of a local mixed-mode I/II/III state along the crack front for a slant or double shear-fracture type. A derived mapping function enables the determination of the fracture type from the surface plastic zone, along with the current crack tip loadings during the test. Finally, a transition of the fracture type also leads to a short-term delay in the crack propagation rate. Based on this information $\mathrm{d}a / \mathrm{d}N- \Delta K_{\mathrm{eqv}}$ curves are computed with regards to the local 3D crack orientation.
\end{abstract}


\end{frontmatter}

\section{Introduction}
Fatigue crack growth is influenced by various factors, including the microstructure of the material, loading conditions, and the environment \cite{Schijve.2003}. According to Ritchie \cite{Ritchie.1999}, microscopic mechanisms drive the crack propagation behaviour in ductile materials. He separates them into intrinsic (damaging) and extrinsic (shielding) fatigue crack growth mechanisms. Understanding those mechanisms and their interactions with microstructural features, the crack tip stress field and environmental conditions is fundamental for developing more accurate crack propagation models relying on a physical foundation \cite{S.Chowdhury.2019}. Such models can improve service life predictions of structural components, leading to a more resource-efficient use of materials \cite{Sagrillo.2022}. 

It is well known in fracture mechanics community that the plastic zone in front of the crack tip represents the crack tip loading condition during fatigue crack growth. Therefore, the plastic zone has received significant attention in research. Finite element (FE) analysis can be used to simulate the size and shape of the plastic zone in the bulk of the material. Recent experimental \cite{Gonzales.2019} and numerical \cite{Benrahou.2007, Subramanya.2005} investigations revealed that its shape differs from the classical dog bone model, proposed by Dugdale \cite{Dugdale.1960}. However, factors such as the specimen thickness, the yield stress, the (multi-axial) loading conditions or the crack front geometry affect the stress singularity of the crack tip and thus the shape of the plastic zone. In contrast to pure mode I loading, mixed-mode loading conditions cause the plastic zone to lose its symmetry feature. Comparing the butterfly shape of the plastic zone can indicate the direction of crack propagation \cite{Golos.2000, Wasiluk.2003}. To evaluate both in-plane and out-of-plane constraints, it is also possible to consider characteristics that classify the size and shape of the plastic zone \cite{Huang.2020}. Recently, Zeinedini \cite{Zeinedini.2019} performed a 3D FE analysis under mixed-mode loading and identified a relationship between plastic zone shape and the loading conditions. He adds that the plastic zone is also asymmetric when a mode III loading is added to a predominant mode I or mode II loading condition.

To simplify the description of the the crack tip stress field including the plastic zone, two-parameter approaches such as the \textit{K-T}-concept, the \textit{J-Q}-concept or the \textit{J-T}-approach can be utilised. An extension towards three-parametric concepts, such as the \textit{K-T-$T_z$}-approach, takes into account out-of-plane and three dimensional effects \cite{She.2007, Wang.2014}. The crack tip constraints describe the stress and strain state at crack front. Variations in these parameters can affect fracture toughness \cite{Pluvinage.2014}. Out-of-plane constraints are related to the specimen's thickness, while in-plane constraints are affected by the crack size, specimen geometry, loading type, and geometry \cite{Betegon.1991}.

Even under pure mode I loading, fatigue crack fronts in aluminium alloys tend to form a shear crack with increasing crack length and stress intensity factor (SIF) \cite{Zuidema.1988, Breitbarth.2020, Felter.2017}. Yamada et al. observed a transition from an initially flat fracture surface towards a double shear (\textit{V-fracture}) or a single shear (\textit{Slant fracture}) as well as transitions from double to single shear \cite{Yamada.2007}. This tilting process often results in a local change in growth rate and therefore has a direct effect on fatigue life \cite{DuPont.2017, Kocanda.2012, DaCarvalhoCunha.2017}. Many numerical studies associate this phenomenon with a local decrease in the crack driving force, assuming that the mode I stress intensity factor is considered to be the crack driving force. This assumption is supported by Bakker \cite{Bakker.1992}, who performed 3D FE simulations with single, double and partially slanted crack front geometries. He confirmed the change from pure mode I loading to mixed-mode loading depending on the crack front geometry. However, the actual cause for crack front geometry changes such as shear lips of slanted crack fronts under mode I loading still remains open \cite{Mahgoub.2003}. Current research focuses on quantifying three-dimensional effects with regards to the crack shape \cite{Camas.2011, BELLETT.2006}, stress state \cite{BELLETT.2005, Kotousov.2013}, three-dimensional crack closure mechanism \cite{Alizadeh.2007,FlorianPaysan.2022} or thickness effects \cite{Liu.2020, Chang.1999}.

In experimental mechanics, digital image correlation (DIC) has become a state-of-the-art tool. By using high resolution DIC (HR-DIC), the butterfly shape of the plastic zone can be identified in real fatigue crack growth experiments. Paysan et al. proposed a new robotic HR-DIC system to automatically capture the evolution of the plastic zone during fatigue crack propagation \cite{Paysan.2023}. The determination of stress intensity factors based on experimentally measured displacement field, has improved over recent years \cite{Hebert.2022}. However, neglecting non-linear effects such as crack branching or crack tip blunting can lead to inconsistent conclusions \cite{Gonzales.2017, Burnett.2015}. Several fatigue crack growth experiments on aluminium alloys have shown that ignoring out-of-plane crack growth can lead to significant underestimations of crack growth rates and overestimation of stress intensity factors \cite{Forth.2003}. This will lead to non-conservative predictions of fatigue life \cite{Forth.2002}.

This paper introduces a new methodology that correlates flat and slant fractures to the shape of the plastic zone at the specimen surface. In fatigue crack growth experiments, the plastic zone is first captured using HR-DIC. Following, the plastic zone at the surface is quantified by its geometric shape and size. Afterwards, the fracture surfaces are  digitised using a optical 3D scanner.  Subsequently, we classify the fracture surface on a macroscopic level as flat, slat, or transition type. Using the experimentally determined fracture surface characteristics, a 3D elastic-plastic finite element simulation is performed to evaluate the relationship between crack front tilting and surface plastic zone shape. This outlines the three-dimensional and through-thickness effects of crack propagation that affect crack propagation rates and therefore fatigue life estimations.

\section{Methods}

\subsection{Material characterization}

\begin{figure}[ht]
    \centering
    \includegraphics[width=7.8 cm]{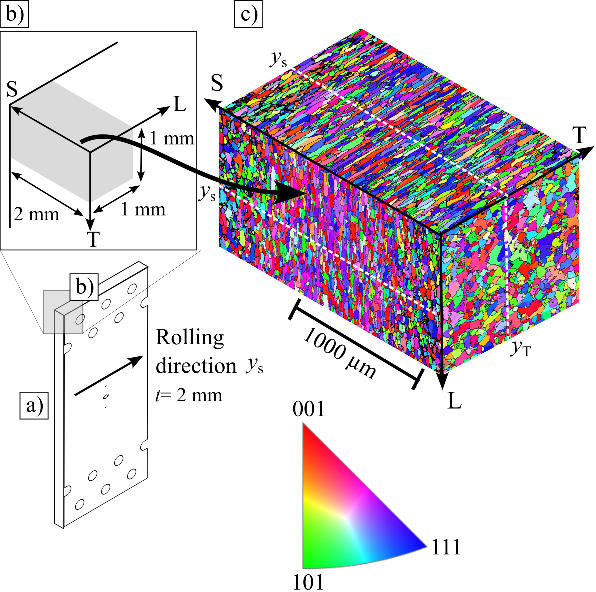}
    \caption{EBSD maps of the rolled AA2024-T3 aluminum sheet material (\textit {t}=2mm) with grain boundaries and IPF map (c). The EBSD sample (b) was extracted from the original sample (a). The coloured representation corresponds to the crystallographic orientation perpendicular to the respective plane}
    \label{fig:EBSD}
\end{figure}

The study used rolled aluminium AA2024-T3 (AlCu4Mg1) sheet material with a thickness of 2mm. The mechanical properties along the rolling direction provided by the supplier are as follows: Yield strength $\sigma_{\mathrm{yield}} =349  \, \mathrm{MPa}$, ultimate tensile strength $R_{\mathrm{m}} =466\, \mathrm{MPa}$, elongation at brake $A = 18 \%$ \cite{AMAGrollingGmbH.2022}. To characterize the grain size distribution as well as the material texture, electron backscatter diffraction (EBSD) scans were performed. All samples were ground and polished up to $3 \, \mathrm{\mu m}$ with diamond suspension, followed by electrolytical polishing for 12 seconds with a current of 24 V in perchloric acid/ethanol. The microstructural analyses were performed using scanning electron microscopy (SEM) (DSM Ultra 55, Carl Zeiss, Germany) equipped with an EBSD detector (Aztec, Oxford Instruments, UK). The EBSD analyses were performed at 20 kV as scanning voltage. To capture the grain structure and orientation over complete sheet thickness, a section of approx. 1 x 2 mm² was set for the T-S and L-S plane, where the L-T-plane section is about 1 x 1 mm² (Figure \ref{fig:EBSD} b). Figure \ref{fig:EBSD} c) shows the EBSD inverse pole figure (IPF) maps and grain boundaries (15° disorientation angle threshold) for all directions L-T, T-S and L-S. The coloured representation corresponds to the crystallographic orientation perpendicular to the respective plane. With regards to the multiple of uniform density (MUD), the pole figure reveals a preferred orientation of the \{100\} plane towards its normal direction (L), as well as an increased and symmetric appearance of the \{111\} and \{101\} plane. This confirms a texture within the sheet material with a preferred orientation generated by the rolling process The texture causes orientation-dependent anisotropic material properties, as has been also observed by other researchers \cite{Chen.2021, Wu.2018}. Due to the machining process, the material shows a grain elongation along the rolling direction L as well as an inhomogeneous grain size distribution along thickness S. Both L-S and T-S direction reveal an increased proportion of elongated grains in the centre of the S-plane. The analysis of grain aspect ratio (Length to height of a single grain) along the sectional planes $y_{\mathrm{S}}$ and $y_{\mathrm{T}}$ marked in Figure \ref{fig:EBSD} c): The L-S and T-S planes show an inhomogeneous grain morphology in which the grain elongation in the center of the sample is clearly pronounced. The grains of the L-T plane show a uniform distribution across the section plane $y_{\mathrm{T}}$, as well as a round, less elliptical appearance with a lower aspect ratio. The grains near the surface are nearly equiaxial whereas there is a strong inhomogeneity with elongated grains towards the rolling direction in the subsurface.

\subsection{Experimental setup}

\begin{figure*}
    \centering
    \includegraphics[width=\textwidth]{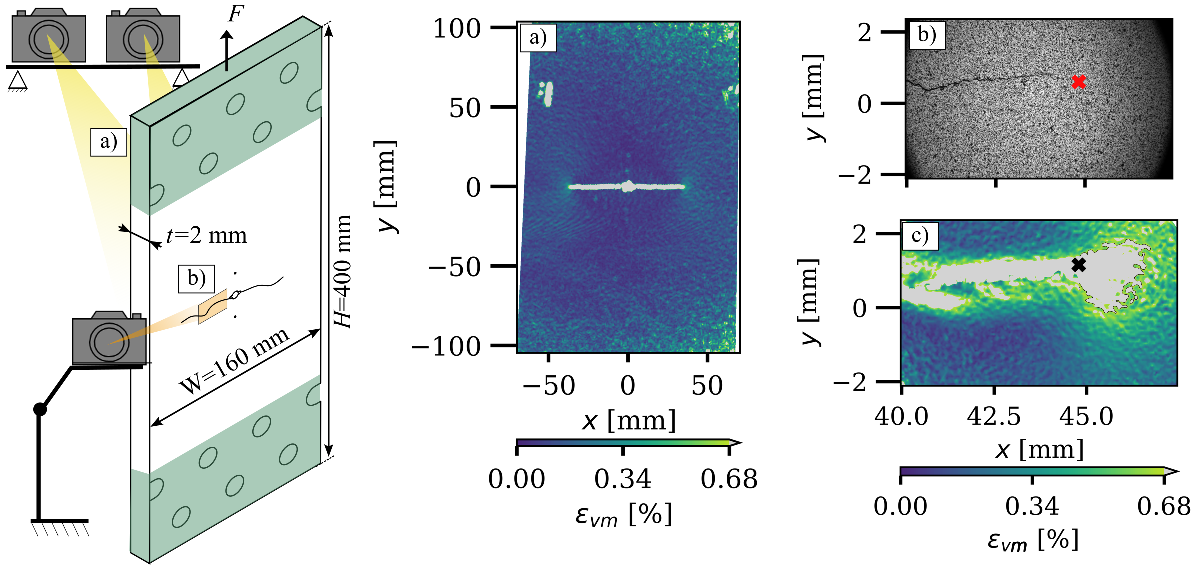}
    \caption{Overview of experimental setup and numerical model: MT-160 specimen with 2D HR-DIC and 3D DIC system, (a) strain field captured by 3D DIC, (b) High resolution microscope image of crack growth, with detected crack tip, (c) strain field captured by 2D HR-DIC}
    \label{fig:Versuchaufbau}
\end{figure*}

The investigated specimens are standardized middle tension (MT) specimen (Width = $160 \, \mathrm{mm}$) in T-L (crack growth parallel to rolling direction) and L-T (crack growth perpendicular to rolling direction) orientation. The crack propagation tests were carried out in accordance to the ASTM E674-13 standard \cite{ASTMInternational.} at a load ratio of $R=0.1$ and a maximum load of $F_{\mathrm{max}}=15\, \mathrm{kN}$. An artificial notch with a length of $2a=16 \, \mathrm{mm}$ was cut into the centre of the specimen. The fatigue crack propagation test was then carried out with a sinusoidal load and constant amplitude at a test frequency of $f=20 \, \mathrm{Hz}$. The crack length was measured with the direct current potential drop (DCPD) method using Johnson's equation \cite{Johnson.1965}. 
DIC was used for full-field displacement measurements on both sample surfaces. A commercial GOM Aramis 12m 3D DIC system was applied to the back surface. With a facet size of 25 x 25 pixels and a facet spacing of 16 pixels, a spatial resolution of $0.72 \, \mathrm{mm}$ per facet was achieved. At the front side, a KUKA LBR iiwa cobot moves a Zeiss STEMI 206C stereo microscope with a Basler a2A5320-23umPRO global shutter CMOS camera (Figure \ref{fig:Versuchaufbau} a)). This HR-DIC system captures a 2D displacement field (Figure \ref{fig:Versuchaufbau} c)) with a  field of view is $10.2 \, \mathrm{mm}$ x $5.7\, \mathrm{mm}$. A refined iron oxide speckled airbrush pattern is applied at the specimen front side. At a magnification of 1.6x, a facet size of 40 x 40 pixels and a facet distance of 30 pixels, a spatial resolution of $0.047 \, \mathrm{mm}$ per facet was achieved. Further details of the test rig can be found in \cite{Paysan.2023}. This multi-scale DIC setup allows sufficient resolution to analyse the plastic zone of the propagating crack, but also to determine the deformation of the entire sample.
During the test, images were taken after $\Delta a=0.5\, \mathrm{mm}$ for the minimum, intermediate and maximum loads. The displacement and strain fields were calculated with the GOM Aramis V2020 software. A trained artificial neural network detects the crack path and the crack tip position in the 2D HR-DIC displacement fields. The details of this technique are described in \cite{Melching.2022, Strohmann.2021}. After the test, the crack propagation curve $\mathrm{d}a/\mathrm{d}N-\Delta K$ was determined in accordance to ASTM E674-13 \cite{ASTMInternational.}.

\subsection{Fracture surface characterization}\label{Fractued surface evaluation}

\begin{figure}[ht]
    \centering
    \includegraphics[width=4 cm]{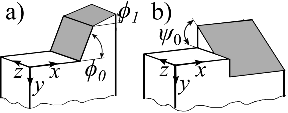}
    \caption{Definition of (a) kink angle $\phi_{0}$ and (b) tilt angle $\psi_{0}$(adapted from \cite{Richard.2014})}
    \label{fig:Winkeldefinitionen}
\end{figure}

\begin{figure}[ht]
    \centering
    \includegraphics[width=7.8 cm]{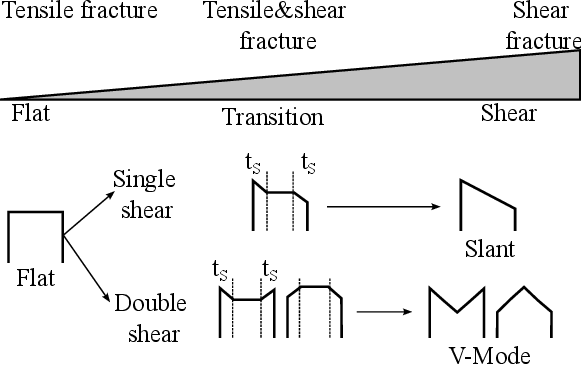}
    \caption{Visualization of classified fracture types with transition from flat to shear fracture according to \cite{Schijve.1981}}
    \label{fig:Modidefinitionen}
\end{figure}

\begin{figure}[ht]
    \centering
    \includegraphics[width=7.8 cm]{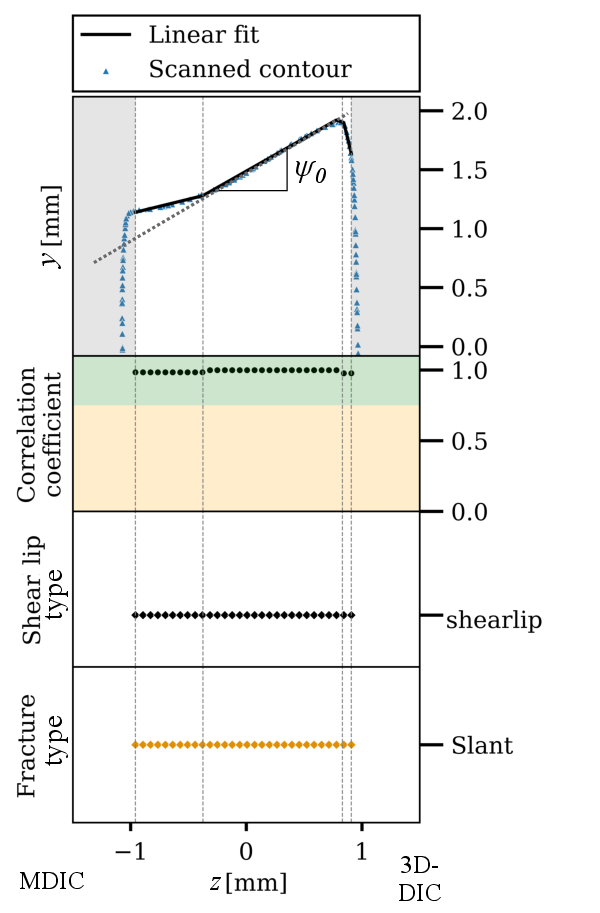}
    \caption{Example of evaluation of slant fracture types of L-T specimen at crack length $a_{x}=$52 mm.}
    \label{fig:Vergleich-V_Slant}
\end{figure}

After testing, the fractured specimens were scanned with a GOM ATOS Q 12M 3D scanner to digitise the topography of the fracture surfaces. The coordinate origin of the z-axis is set to the centre of the specimen and aligned with the coordinate system of the DIC system. The fracture surface was then characterised by local shape descriptors. Figure \ref{fig:Winkeldefinitionen} shows the definitions of the crack kink angle $\phi_{0}$ (a) and the crack tilt angle $\psi_{0}$ (b), around the z-axis and x-axis. To determine the crack tilt angle $\psi_{0}$, y-z cutting planes are set along the x-axis with a distance of $0.1 \, \mathrm{mm}$. To include only the fractured surface, the left and the right 10\% of the total contour of the y-z plane are masked (Figure \ref{fig:Vergleich-V_Slant}, grey marked area). The remaining contour is approximated piece by piece linearly with three segments \cite{JekelCharlesF.VenterGerhard.2019}. To assess the quality of the fit, a correlation coefficient is determined for each linear curve segment and the original contour points. Furthermore, the tilt angle $\psi_{0}$ of the central segment, the total percentage of shear lips width and the linear fit correlation coefficient are used to determine the fracture type. Significant differences in the slope of the surface segments with the centre are associated with shear lips \cite{Ling.1990}. Shear lips are identified by means of case differentiation, taking into account the segment position, the segment width and the angular difference between the segments. Contours with a correlation coefficient $<$ 0.75 for over 50\% of the analysed contour width are classified as a transition type as the linear fit is insufficient due to the contour complexity. Figure \ref{fig:Modidefinitionen} shows the identification of four fracture types, including flat, transition, slant, and V-fracture. To simplify the classification, the cases categorised as \textit{Flat-to-Slant} and \textit{Flat-to-V-fracture} were referred to as Slant or V-fracture. For clarity, the double shear fracture is labelled here as a V-fracture.

\subsection{Numerical analysis} \label{FE-Modellaufbau}

\begin{figure*}
    \centering
    \includegraphics[width=\textwidth]{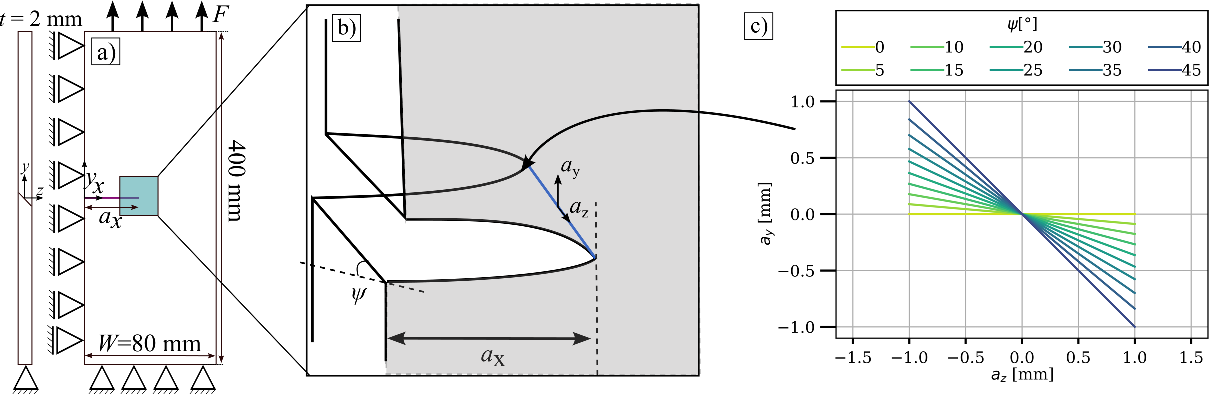}
    \caption{a) Numerical 3D model with boundary conditions, b) Detail view of modeled crack front depending on crack front position and tilt angle, c) Visualization of crack front depending on tilt angle $\psi$}
    \label{fig:FE-Modell}
\end{figure*}

Elastic-plastic 3D FE analyses were performed to analyse the characteristic plastic zone on the specimen surface as a function of the tilt angle $\psi_{0}$ and the ratio of crack length to specimen width \textit {a/W}.
Linear elastic simulations were performed to outline the dependence of tilt angle $\psi_{0}$ towards stress intensity factors $K_{\mathrm{I}}, K_{\mathrm{II}}, K_{\mathrm{III}}$. A total number of 70 simulations were carried out for each parameter study and 65 simulations for comparison with the experiments. The simulations were performed with ANSYS Mechanical APDL Version 2022 R1 on a RedHat Linux Workstation with two Intel Xeon Gold 6240 18C CPUs and a total DDR4-2933 RAM of 256 GB.

Material properties of the AA2024-T3 aluminium alloy are taken from literature \cite{Tamarin.2002}. Young’s modulus and Poisson ratio were set to $E=73.1$ GPa and $\mu=0.33$. Hardening modulus and yield stress were set to $m_{t}=984$ MPa and $R_{p,0.2}=345$ MPa, using a bilinear kinematic hardening model (APDL: BISO) for elastic-plastic simulations. This model neglects anisotropic material properties as well as cyclic material behaviour. The MT160 model was meshed using free mesh and tetrahedral SOLID187 elements. To reduce the complexity of the model, symmetry is used in the centre of the model (APDL: D, ALL, UX, 0). The MT160 specimen is therefore modelled as a half model with a width $W/2=80 \, \mathrm{mm}$. According to the experimental setup, the lower surface was clamped in all directions (APDL: D, ALL, ALL, 0) and a constant load was applied at the top surface (APDL: SFA, ALL, PRES, $\sigma$). This ensures an uniformly distributed load in the normal direction of the surface, which represents the clamped upper side of the test setup (Figure \ref{fig:Versuchaufbau}, green area). The overall mesh can be subdivided in a global mesh with an element size of 1 mm and a refined area around the crack tip (APDL: NREFINE) with an element size of 0.042 mm for the linear-elastic simulation. For the elastic-plastic simulation the domain of mesh refinement was set by estimating the plastic zone size according to Irwin’s formula \cite{Irwin.1957} (under plane stress conditions) for $a/W=0.8$ and a $K_{\mathrm{I}}=50\,\mathrm{MPa\sqrt{m}}$, leading to a refinement area of $r_\mathrm{p}=3.2\,\mathrm{mm}$ along the crack tip. This leads to a refined element size of $0.037\,\mathrm{mm}$ and a model with a total of 1.079.885 nodes. Table \ref{tab:FE-Parameterstudie} lists the parameters for evaluating the dependency of the crack front tilt angle $\psi$ on plastic zone geometry. A total number of 70 simulations are used for the parametric study. 

\begin{figure}[ht]
    \centering
    \includegraphics[width=7.8 cm]{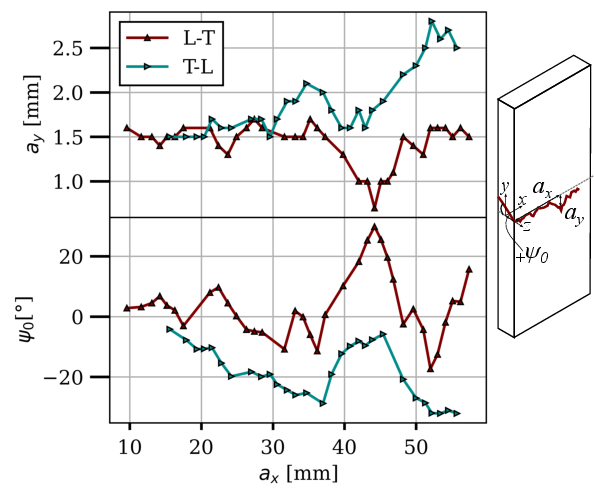}
    \caption{Experimentally determined input parameter crack path $a_{x}$, $a_{y}$ and tilt angle $\psi_{0}$ for elastic-plastic simulation of the plastic zone of L-T and T-L specimen}
    \label{fig:RB_Sim_Exp}
\end{figure}

In order to compare the experimentally determined plastic zone with the numerical model, the parameters given in Figure \ref{fig:RB_Sim_Exp} were used for an additional study. Therefore, the experimentally determined crack path coordinates $a_{x}$ and $a_{y}$ and the tilt angle $\psi_{0}$ of the fracture surfaces are applied to the model. A total of 36 simulations were carried out for the L-T specimen and 29 simulations for the T-L specimen in order to compare them with the experimental investigations of the plastic zone.

The stress intensity factors $K_{\mathrm{I,II,III}}$ are calculated for each node of the crack front by using the volume interaction integral provided in ANSYS (APDL: CINT). This method returns a mean value of six integration paths for each node along the crack front. The node-individual stress intensity factors are then averaged over the crack front.

\begin{table} [t] 
    \centering
	\begin{tabular}{l|ll}
	    Parameter & Variations  \\ \hline
	    \makecell[l]{Length to width \\ ratio $a/W$}  &  \makecell[l]{ $a/W=$[0.3, 0.4, 0.5, 0.6, \\ 0.7, 0.8]}   \\ 
	    \makecell[l]{Crack tip $a_{x}$}   &\makecell[l]{ $a_{x}=$[16, 24, 32, 40, 48,\\ 56, 64] mm}  \\
	    \makecell[l]{tilt angle $\psi$}  & \makecell[l]{$\psi=$[0, 5, 10, 15, 20, \\25, 30, 35, 40, 45] \degree} \
	\end{tabular}
	\caption{Parameter for elastic-plastic parameter study of the influence of tilt angle towards shape of plastic zone}
	\label{tab:FE-Parameterstudie}
\end{table}

\subsection{Plastic zone evaluation}\label{Plastic zone evaluation}

\begin{figure}[ht]
    \centering
    \includegraphics[width=7.8 cm]{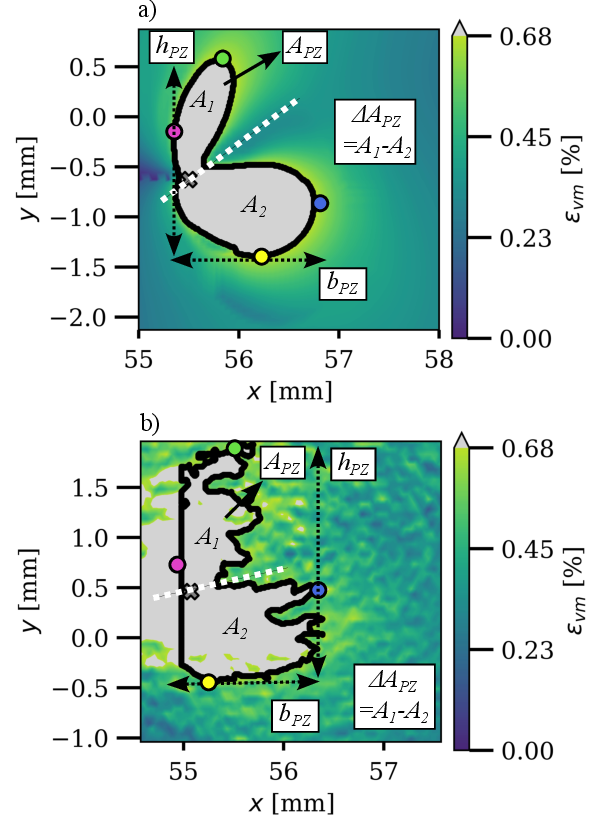}
    \caption{a) Numerically determined plastic zone at the surface strain field with dimension points and definitions of height $h_{\mathrm{pz}}$, length $b_{\mathrm{pz}}$ and division line between upper area $A_1$ and lower area $A_2$, b) experimentally determined plastic zone in HR-DIC data with same; Example here: T-L, $a_{x}=55.1$ mm, $a_{y}=0.37$ mm, $\Delta K_{\mathrm{I,ASTM}}=25.1\,\mathrm{MPa\sqrt{m}}$, $\Delta K_{\mathrm{I,corr}}=21.8\,\mathrm{MPa\sqrt{m}}$, $\Delta K_{\mathrm{II,corr}}=-0.009\,\mathrm{MPa\sqrt{m}}$, $\Delta K_{\mathrm{III,corr}}=-13.2\,\mathrm{MPa\sqrt{m}}$, $\psi_{0}=-31.10\degree$, $\phi_{0}=-0.9 \degree$.}
    \label{fig:PZ_Evaluation}
\end{figure}

For direct comparison, the contour of the plastic zone was analysed both for HR-DIC displacement data and for the free surface of the 3D finite element model. The analyses of the plastic zone always refer to the displacement field at maximum load. See Figure \ref{fig:PZ_Evaluation} a) for experimental input and b) for input from finite element analysis. To evaluate the plastic zone at the surface of the FE model, only the nodes at the surface position of $z=1$ mm are selected (see Figure \ref{fig:FE-Modell} b), grey marked area), whereas the input from the HR-DIC data is already given as a 2D-displacement field. In the first step, the crack tip position is automatically determined in the HR-DIC displacement data, whereas the position is given as input in the FE-Analysis. Within the displacement field, the contour of the plastic zone is set by masking the linear-elastic area, where the total strain $\varepsilon_{\mathrm{total}}$ is lower than the von Mises equivalent strain $\varepsilon_{\mathrm{vm}}$ of 0.68 \%. This is an uniaxial estimation of the yield criterion given by:
\begin{equation}
\varepsilon_{\text {total }}=\varepsilon_{\mathrm{el}}+\varepsilon_{\mathrm{pl}}=\frac{R_{\mathrm{p} 0,2}}{E}+\varepsilon_{\mathrm{pl}}
\end{equation}

with $\varepsilon_{\mathrm{pl}}=0.2$ according to the definition of $R_{\mathrm{p} 0,2}$.  

Also, the plastic wake behind the crack tip is masked for the experimental data. This is not necessary for the FE model as a quasi-static simulation without crack propagation does not create a plastic wake. In the case of the experimental data, even secondary cracks can be detected. This is the case if more than one plastic zone is recognised at the crack tip.
Due to sensitivity of the optical sensors, air movement, varying lighting conditions, reflections, the speckle pattern quality or out-of-plane motion, DIC images are subjected to inherent noise that affect the DIC evaluation  \cite{Lecompte.2006, Zappa.2014, Sutton.2008}. Therefore, several contours can also be detected without a recognisable wing shape. In this case, we consider those as artefacts and they were not taken into account for further evaluation.
Since the FE results are not influenced by scattering or secondary cracks, the plastic zone is always a smooth consistent contour. The contour of the plastic zone $A_{\mathrm{pz}}$ envelops the masked area (see black contours in Figure \ref{fig:PZ_Evaluation} a) and b)). The difference between upper wing $A_1$ and lower wing $A_2$, being separated by a line between crack tip and minimum contour extension, defines the area difference $\Delta A_{\mathrm{pz}}$. The height $h_{\mathrm{pz}}$ is defined as horizontal distance between minimum and maximum horizontal extension point. The length $b_{\mathrm{pz}}$ is defined as vertical distance between minimum and maximum vertical extension point (Figure \ref{fig:PZ_Evaluation}). 

\section{Results}

The results of this study encompass several aspects. Firstly, the influence of tilt angle on stress intensity factors $K_{\mathrm{I}}$, $K_{\mathrm{II}}$, $K_{\mathrm{III}}$, and the equivalent stress intensity factor $K_{\mathrm{eqv}}$ is examined. Secondly, the geometry parameters of the plastic zone on the surface are affected by the stress intensity factor $K_{\mathrm{I}}$ and the tilt angle $\psi$, as grounded in a numerical parameter study. Finally, these findings are utilized to explore the differing crack propagation rates of the L-T and T-L orientations based on three-dimensional analysis.

\subsection{Parameter study: Influence of tilt angle on stress intensity factors $K_{\mathrm{I}}$, $K_{\mathrm{II}}$, $K_{\mathrm{III}}$, $K_{\mathrm{eqv}}$}\label{SIF_Tilting}

\begin{figure}[ht]
    \centering
    \includegraphics[width=7.8 cm]{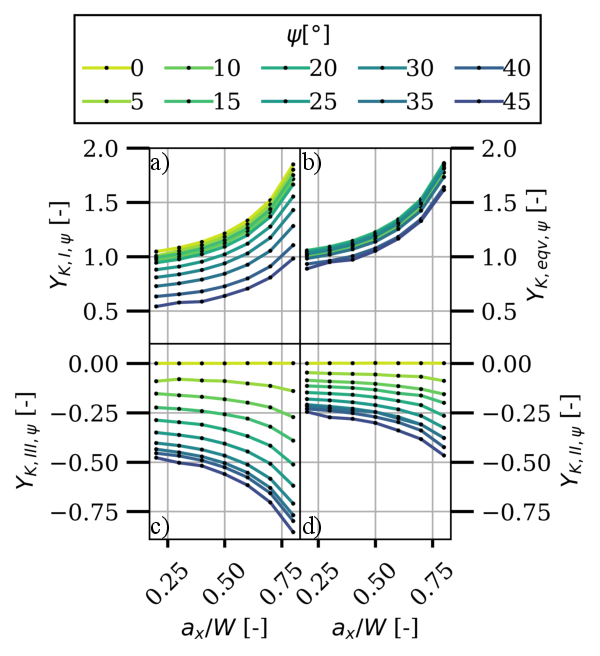}
    \caption{Normalized stress intensity factors a) mode I $Y_{\mathrm{I}}$, b) mode II $Y_{\mathrm{II}}$, c) mode III $Y_{\mathrm{III}}$, d) equivalent stress intensity factor $Y_{\mathrm{eqv}}$ depending on tilt angle $\psi$}
    \label{fig:Geometriefaktor}
\end{figure}

Figure \ref{fig:Geometriefaktor} outlines the dependency of the tilt angle $\psi$ and normalized crack lenght $a_{x}/W$ towards normalized stress intensity factors $Y_{\mathrm{I},\psi}, Y_{\mathrm{II}\psi}, Y_{\mathrm{III}\psi}, Y_{\mathrm{eqv}\psi}$. To calculate the equivalent stress intensity factor $K_{\mathrm{eqv}}$, Richard's formula \cite{Richard.2014} is used. 
\begin{equation}
K_{\mathrm{eqv}}=\frac{K_{\mathrm{I}}}{2}+\frac{1}{2} \sqrt{K_1^2+4 \cdot K_{\mathrm{II}}^2+5.336 \cdot K_{\mathrm{III}}^2}
\label{Keqv}
\end{equation}

The geometry factors corresponding to the nominal load are normalized using the equation

\begin{equation}
K_{i,\psi}=\frac{F}{W \cdot t} \cdot \sqrt{\pi \cdot a} \cdot Y_{i, \psi}
\end{equation}

where $i=[\mathrm{I,II,III,eqv}]$.

As the tilt angle $\psi$ increases, $Y_{\mathrm{I},\psi}$ decreases up to 40 \%. Crack tilting also leads to a mixed-mode-loading condition as evidenced in the increase of $Y_{\mathrm{II},\psi}$ and $Y_{\mathrm{III},\psi}$. This also impacts the mixed-mode equivalent stress intensity factor $Y_{\mathrm{eqv},\psi}$, which decreases by 20\% compared to the decrease in $Y_{\mathrm{I},\psi}$.

\begin{figure}[ht]
    \centering
    \includegraphics[width=7.8 cm]{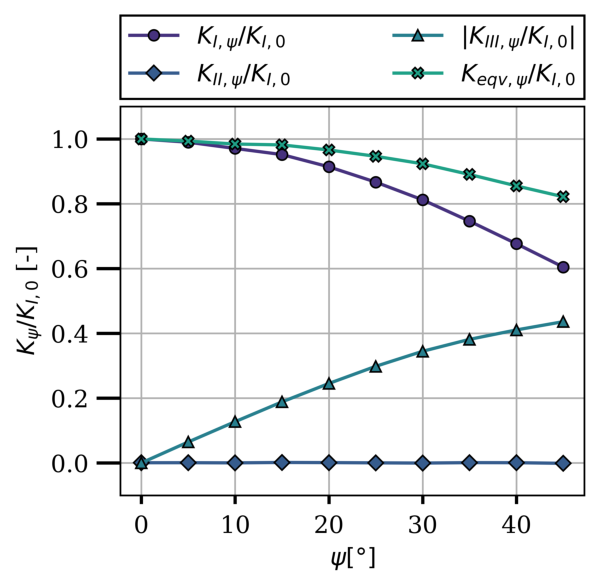}
    \caption{Normalized stress intensity factors $K_{i,\psi}$}
    \label{fig:Abnahme_K}
\end{figure}

Figure \ref{fig:Abnahme_K} shows the stress intensity factors $K_{i,\psi}$ normalized towards the stress intensity factor of a straight crack $K_{\mathrm{I,0}}$ as a function of $\psi$. This relationship can be used as a correcting function to describe the distribution of $K_{i,\psi,corr}$ when the crack tilt angle $\psi$ and the stress intensity factor for a straight crack $K_{\mathrm{I,0}}$ are known. By utilizing a quadratic fitting function, $K_{i,\psi, corr}$ can be calculated as follows:

\begin{equation}
K_{i,\psi, \mathrm{corr}}=(a_{i}\cdot\psi^{2}+b_{i}\cdot\psi+c_{i})\cdot K_{\mathrm{I,0}}
\end{equation}

\begin{table} [t] 
    \centering
    \begin{tabular}{l|l|l|l}
	    SIF & \makecell[l]{$a_{i}\cdot10^{-4}$} & \makecell[l]{$b_{i}\cdot10^{-4}$} & \makecell[l]{$c_{i}\cdot10^{-4}$} \\ \hline
        $K_{\mathrm{I,}\psi,\mathrm{corr}}$ & $-1.77$ & $-9.14$ &  10000 \\
        $K_{\mathrm{II,}\psi,\mathrm{corr}}$ & $-0.015$  & $0.445$  & $4.42$  \\
        $K_{\mathrm{III,}\psi,\mathrm{corr}}$ & $-1.08$  & $147$ & $-47.7$ \\
        $K_{\mathrm{eqv,}\psi,\mathrm{corr}}$ & $-0.928$  & $2.51$ & 9966\
	\end{tabular}
	\caption{Parameters for fitting function to calculate $K_{i,\psi}$ based on a given tilt angle  $\psi$ and stress intensity factor for a straight crack $K_{\mathrm{I,0}}$}
	\label{tab:Korrekturfunktion}
\end{table}

\subsection{Parameter study: Influence of tilt angle and load on the shape and size of the plastic zone} \label{Parameterstudy}

\begin{figure}[ht]
    \centering
    \includegraphics[width=7.8 cm]{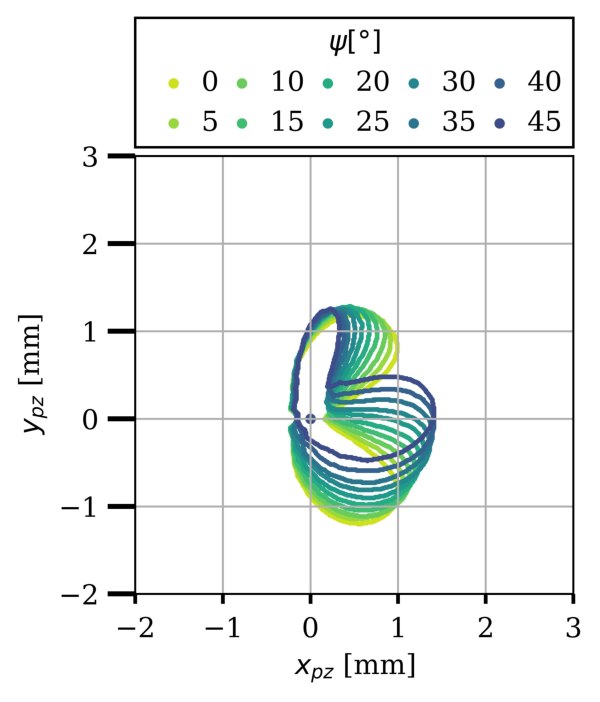}
    \caption{Contour of the plastic zone at $K_{\mathrm{I}}=29.9\,\mathrm{MPa\sqrt{m}}$ with different crack front tilt angle $\psi$}
    \label{fig:Konturvergleich_Winkel}
\end{figure}

\begin{figure}[h]
    \centering
    \includegraphics[width=7.8 cm]{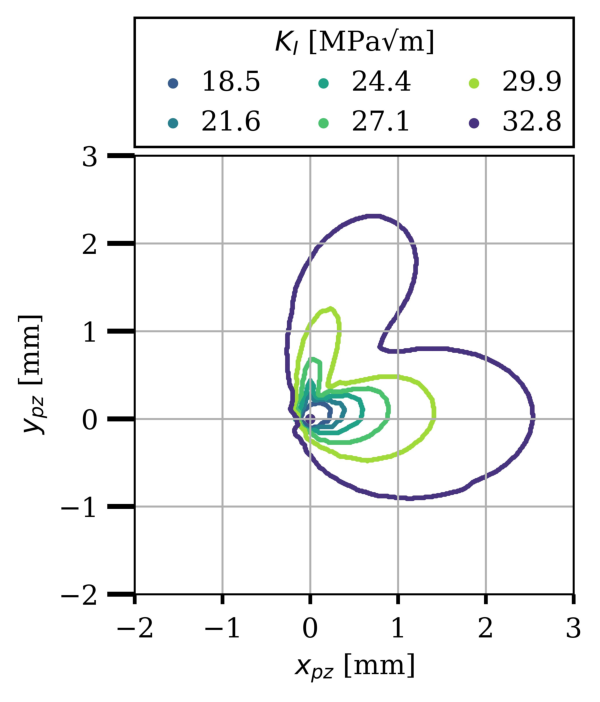}
    \caption{Contour of the plastic zone at crack front tilt angle $\psi=35\degree$ depending on $K_{\mathrm{I}}$}
    \label{fig:Konturvergleich_SIFS}
\end{figure}

\begin{figure}[ht]
    \centering
    \includegraphics[width=7.8 cm]{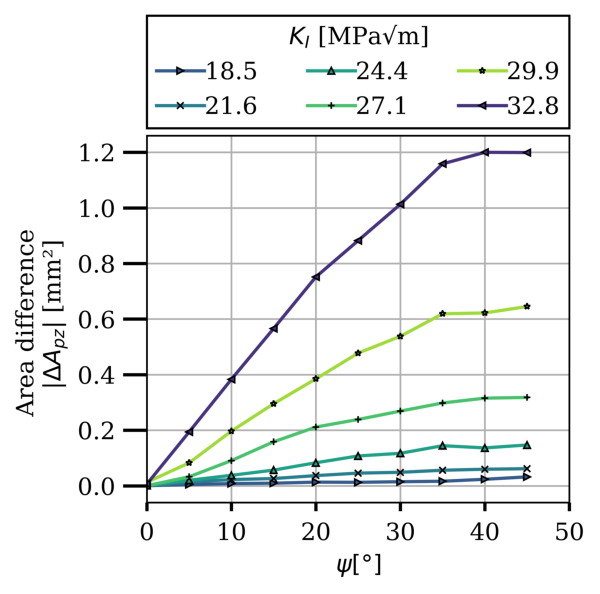}
    \caption{Relationship between crack front tilt angle $\psi$ and area difference of the plastic zone wings $\lvert\Delta A_{\mathrm{pz}}\rvert$ at different $K_{\mathrm{I}}$}
    \label{fig:Flächendifferenz}
\end{figure}

\begin{figure}[ht]
    \centering
    \includegraphics[width=7.8 cm]{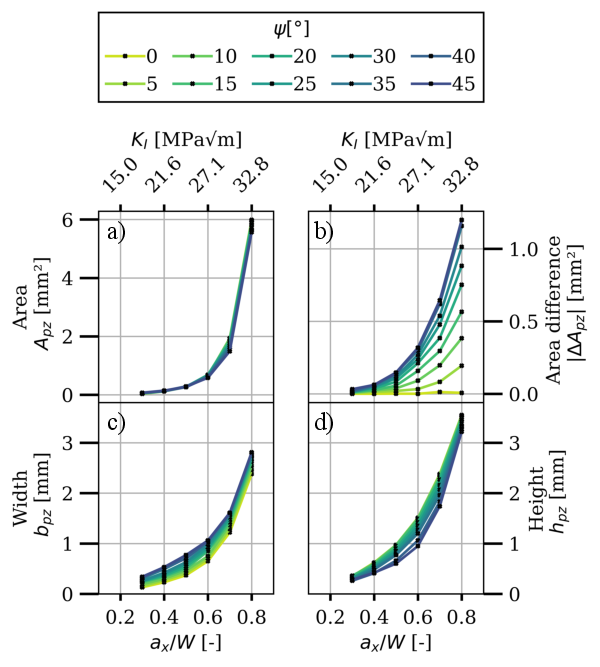}
    \caption{Characterization of the plastic zone: a) area $A_{\mathrm{pz}}$, b) absolute wing area difference $\Delta A_{\mathrm{pz}}$, c) width $b_{\mathrm{pz}}$, d) height $h_{\mathrm{pz}}$}
    \label{fig:PZ_Sim_Ergebnisse}
\end{figure}

Figure \ref{fig:Konturvergleich_Winkel} compares the contour of simulated plastic zones for a nominal $K_{\mathrm{I}}=30\,\mathrm{MPa\sqrt{m}}$ and variation of the tilt angle $\psi_{0}$ for the parameter study of the FE model described in section \ref{FE-Modellaufbau}. The contour is captured at the specimens surface $z \, = \, 1 \, \text{mm}$ (see Figure \ref{fig:FE-Modell} b) grey marked area). The stress intensity factor $K_{\mathrm{I}}$ was calculated according to \cite{ASTMInternational.} neglecting the tilted crack front. With increasing tilt angle $\psi_{0}$ the contour evolves more and more asymmetric. The shape also differs from the classical and symmetric dog-bone model \cite{Dugdale.1960}. 
As introduced in section \ref{Plastic zone evaluation}, Figure \ref{fig:PZ_Evaluation} illustrates the dependency of the geometry parameters, namely (a) area $A_{\mathrm{pz}}$, (b) wing area difference $\lvert\Delta A_{\mathrm{pz}}\rvert$, (c) width $b_{\mathrm{pz}}$ and (d) height $h_{\mathrm{pz}}$. 
The area $A_{\mathrm{pz}}$ shows a quadratic relationship along the stress intensity factor $K_{\mathrm{I}}$ that reflects the power law proposed by Irwin \cite{Irwin.1957}. The area $A_{\mathrm{pz}}$ seems to be independent from the tilt angle $\psi_{0}$. This can be explained by a constant energy input and an angle-independent plastic energy dissipation within the plastic zone \cite{Li.2003}. Width $b_{\mathrm{pz}}$ and height $h_{\mathrm{pz}}$ also show a quadratic relationship to $K_{\mathrm{I}}$, but also a dependency from tilt angle $\psi_{0}$. When the tilt angle increases, the shape of the plastic zone tilts along the x-axis. This leads to a reduction in height around 10 \% for each increasing angle increment. The lower wing expands, leading to an increased width $b_{\mathrm{pz}}$. This characteristics is dependent on $K_{\mathrm{I}}$ and the tilt angle $\psi_{0}$. This is also evident from the absolute area difference between the upper and lower wing $\lvert\Delta A_{\mathrm{pz}}\rvert$, which characterises the asymmetry of the shape. An increasing tilt angle $\psi$ leads to an increased width $b_{\mathrm{pz}}$ which results in an asymmetric area ratio $\lvert\Delta A_{\mathrm{pz}}\rvert$ of the upper and lower wing. This relationship is outlined in Figure \ref{fig:Flächendifferenz}. A linear relationship between area difference $\lvert\Delta A_{\mathrm{pz}}\rvert$, and tilt angle $\psi$ can be identified.

\subsection{Crack propagation curve $\mathrm{d}a/\mathrm{d}N-\Delta K_{\mathrm{I}}$}\label{Rifokurve}

\begin{figure}[ht]
    \centering
    \includegraphics[width=7.8 cm]{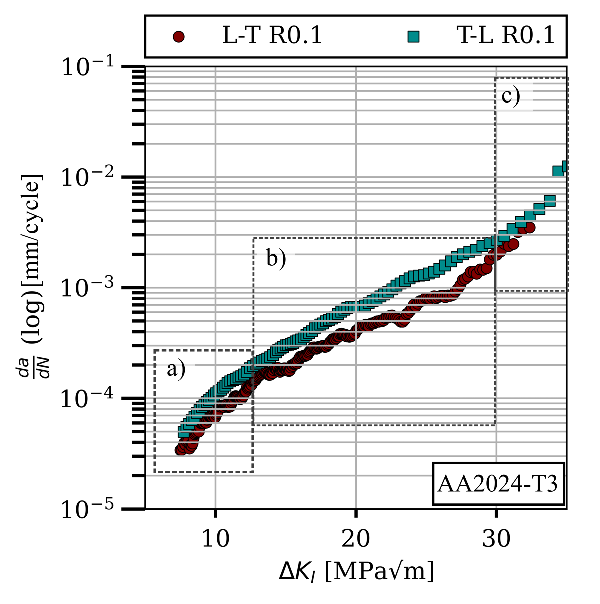}
    \caption{Comparison of crack propagation curve $\mathrm{d}a/\mathrm{d}N-\Delta K_{\mathrm{I}}$ of L-T and T-L oriented specimen with categorization in three sub-areas}
    \label{fig:Rifokurve}
\end{figure}

Figure \ref{fig:Rifokurve} compares the crack propagation curves $\mathrm{d}a/\mathrm{d}N$ vs. $\Delta K_{\mathrm{I}}$ for both tested crack orientations L-T and T-L evaluated according to ASTM E674-13. Three distinct sections can be identified:

In section (a), the crack propagation behavior of L-T and T-L orientations is comparable in terms of propagation rate and stress intensity factor.

Section (b) begins at $\Delta K_{\mathrm{I}}>13 \, \mathrm{MPa\sqrt{m}}$ and a crack propagation rate of $\mathrm{d}a/\mathrm{d}N>2\cdot10^{-4} \, \mathrm{mm/cycle}$. Here, the crack in the T-L oriented specimen propagates faster for an identical $\Delta K_{\mathrm{I}}$. Additionally, the crack propagation curve of the L-T specimen exhibits fluctuations in the crack propagation rate, while the T-L curve is smoother.

For $\Delta K_{\mathrm{I}}>30 \, \mathrm{MPa\sqrt{m}}$ and $\mathrm{d}a/d\mathrm{N}>2\cdot10^{-4} \, \mathrm{mm/cycle}$, section (c) begins. Here, both curve progressions equalize towards a similar accelerating crack propagation rate.

\subsection{Characterization of the three-dimensional crack propagation behavior based on fracture surface analysis}

\begin{figure}[ht]
    \centering
    \includegraphics[width=7.8 cm]{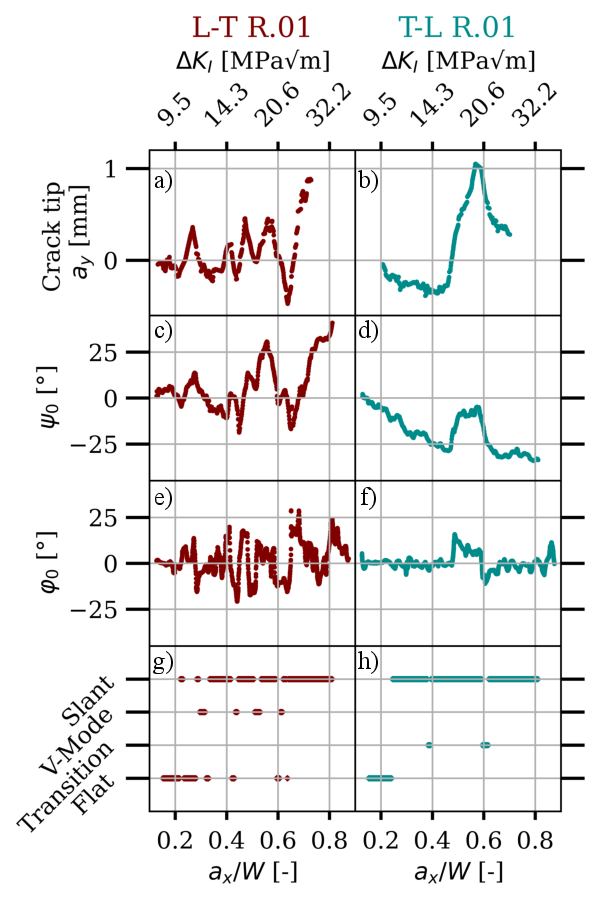}
    \caption{Comparison of the experimentally determined parameters from HR-DIC evaluation and fracture surface characterization, a,b) microscopical crack path, c,d) tilt angle $\psi_{0}$, e,f) kink angle $\phi_{0}$, g,h) classified fracture types}
    \label{fig:Vergleich-LT_TL}
\end{figure}

\begin{figure*}
    \centering
    \includegraphics[width=\textwidth]{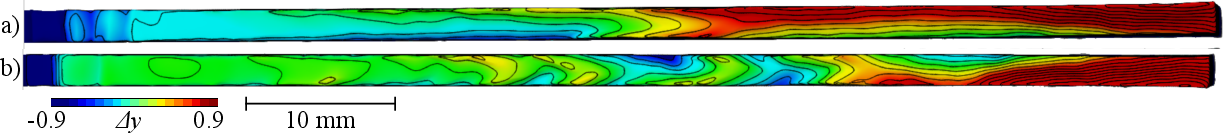}
    \caption{Top view and comparison of the fracture surface captured via 3D Scan with height difference $\Delta y$ towards initial crack plane, a) T-L specimen, b) L-T specimen}
    \label{fig:Bruchflächen}
\end{figure*}

Figure \ref{fig:Bruchflächen} visualizes the fracture surfaces of T-L and L-T specimen captured via 3D scan. Using the method described in \ref{Fractued surface evaluation}, it is possible characterize the three dimensional characteristics. Figure \ref{fig:Vergleich-LT_TL} compares the crack path from HR-DIC evaluation (a,b), tilt angle $\psi_{0}$ (c,d), kink angle $\phi_{0}$ (e,f) as well as the classified fracture types (g,h). The stress intensity factor $\Delta K_{\mathrm{I}}$  was calculated according to \cite{ASTMInternational.} with regards to the specimen geometry for the experimentally given parameters load ratio $R=0.1$, maximum load $F_{\mathrm{max}}=15\,\mathrm{kN}$ and the crack length $a$ evaluated via DCPD. As the size of the plastic zone exceeds the microscopic field of view for $\Delta K_{\mathrm{I}}>25\mathrm{MPa\sqrt{m}}$ and $a_{x}/W>0.7$, data points of the macroscopic crack path in Figure \ref{fig:Vergleich-LT_TL} a) and b) are missing. Regarding the crack path, the L-T orientated specimen shows an irregular crack path with several turning points in a stress intensity factor range between 11 and 25 $\mathrm{MPa\sqrt{m}}$. This behaviour is also evident in the fracture pattern, where periodic crack kink and tilt can be observed. This also results in multiple tilting of the crack front leading to multiple changes of the fracture types up to a range of 25 $\mathrm{MPa\sqrt{m}}$. Regarding the T-L orientated specimen, the crack path remains fairly stable and straight, being characterized by lower kink angles. The crack front deflects with increasing crack length. The crack path changes its direction only twice at 16 $\mathrm{MPa\sqrt{m}}$ and 19 $\mathrm{MPa\sqrt{m}}$ which is accompanied with crack kinking and a change in the tilt direction. This leads to crack fronts with less changes in fracture types compared to L-T. Changes in crack path direction correspond to reversals of the tilt angle $\psi_{0}$, partly leading to changes in fracture type, as the tilt angle is a criterion for identifying a slant fracture.

\section{Discussion}

\subsection{Effect of tilting on stress intensity factors}
\begin{figure}[ht]
    \centering
    \includegraphics[width=7.8 cm]{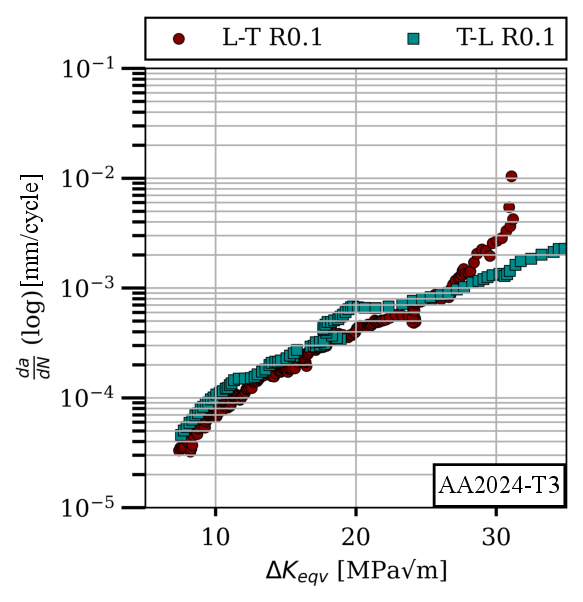}
    \caption{Comparison of the crack propagation curve $\mathrm{d}a/\mathrm{d}N-\Delta K_{\mathrm{eqv}}$ of L-T and T-L oriented specimens, taking into account the local tilt angle in the SIF calculations.}
    \label{fig:Rifokurve_KEQV}
\end{figure}

By using the stress intensity factor $\Delta K_{\mathrm{I}}$ calculated in section \ref{Rifokurve} and the experimentally determined crack tilt angle $\psi_{0}$ as input, the equations outlined in section \ref{SIF_Tilting} can be applied. Figure \ref{fig:Rifokurve_KEQV} illustrates the correlation between the crack growth rate $\mathrm{d}a/\mathrm{d}N$ and the equivalent stress intensity factor $\Delta K_{\mathrm{eqv}}$, which considers mixed-mode loading conditions. While the crack propagation curve based on ASTM E647-15 in figure \ref{fig:Rifokurve} is determined assuming a pure mode I loading, the calculation of $\Delta K_{\mathrm{eqv}}$ takes into account $\Delta K_{\mathrm{III}}$ due to slant crack growth and therefore, mixed-mode crack propagation behaviour. Correlating $\Delta K_{\mathrm{eqv}}$ with the crack growth rate allows a more detailed description of the real crack tip stress field under mixed-mode condition. This correction is particularly important in the  $\Delta K$ regime between $10 \, \mathrm{MPa\sqrt{m}}$ and $28 \, \mathrm{MPa\sqrt{m}}$, where a more accurate stress intensity factor calculation leads to almost superimposed crack propagation curves in the L-T and T-L directions. The deviation for higher $\Delta K_{\mathrm{eqv}}>$ 28 $\mathrm{MPa\sqrt{m}}$ is possibly due to plasticity \cite{Huang.2009} and the proximity to the edge of the sample.

\subsection{Effect of kinking and tilting on crack propagation and fracture surface}

The numerical and experimental results support the results from the literature that crack kinking and crack tilting cause a significant deviation of the stress field at the crack tip from a pure mode I crack with a straight crack front. A slanted or V-shaped crack leads to localised crack growth in mixed-mode \cite{Spagnoli.2013},\cite{Shoemaker.2023}. The effect of the crack orientation on the crack path morphology and therefore, on the  crack growth rate was also outlined by several researcher \cite{Venkateswararao.1988,Maduro.2011}. By comparing the crack propagation curve with the kink angle $\phi_{0}$ and the tilt angle $\psi_{0}$, all curves show a periodic behaviour in the range of $13<\Delta K_{\mathrm{I}}<30 \mathrm{MPa\sqrt{m}}$. This goes along with several changes of fracture types within this range.

\subsection{Experimentally and numerically determined plastic zone}

\begin{figure}[ht]
    \centering
    \includegraphics[width=7.8 cm]{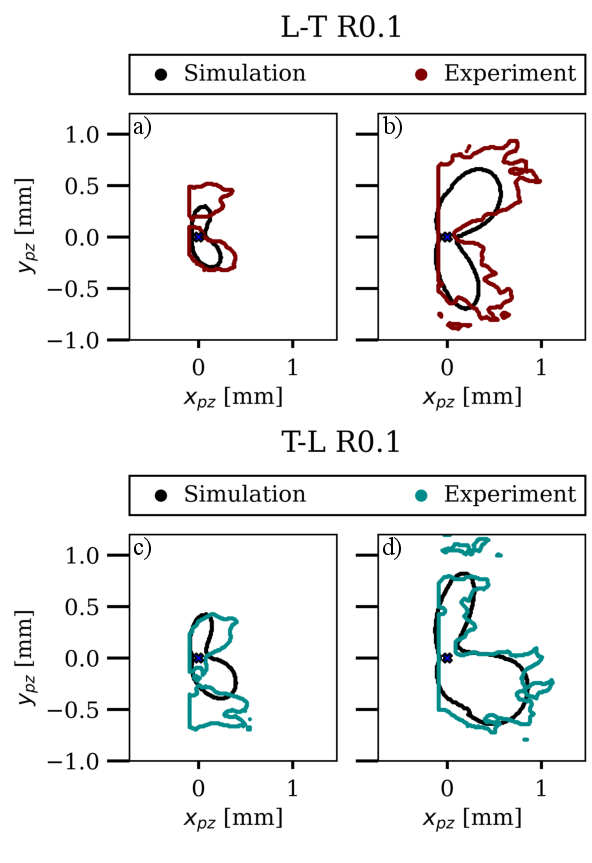}
    \caption{Comparison of experimentally and numerically determined plastic zones according to parameters given in Figure \ref{fig:RB_Sim_Exp}}
    \label{fig:Konturvergleich_PZ}
\end{figure}

\begin{table*} [t] 
    \centering
	\begin{tabular}{l|l|l|l|l|l|l|l}
	    No. & Orientation & \makecell[l]{$\psi_{0}$\\$[\degree]$}&
	    \makecell[l]{$\phi_{0}$\\$[\degree]$}&
	    \makecell[l]{$\lvert\Delta A_{\mathrm{pz}}\rvert$\\\,$\mathrm{[mm^{2}]}$} & \makecell[l]{$\Delta K_{\mathrm{I},\psi,\mathrm{corr}}$\,\\$\mathrm{[MPa\sqrt{m}]}$}& 
	    \makecell[l]{$\Delta K_{\mathrm{II},\psi,\mathrm{corr}}$\,\\$\mathrm{[MPa\sqrt{m}]}$}&
	    \makecell[l]{$\Delta K_{\mathrm{III},\psi,\mathrm{corr}}$\,\\$\mathrm{[MPa\sqrt{m}]}$} \\\hline
a) & L-T R0.1 & -9.78 & 4.4 & - & 14.1 & 0& -1.0 \\
b) & L-T R0.1  & 1.9 & -11.5 & 0.08 & 20.2 & 0 & 0.6 \\
c) & T-L R0.1  & -24 & 0  & - & 15.3 & 0 & -6.5 \\
d) & T-L R0.1  & -27.0 & -4.2 & 0.33 & 19.3 & 0 &- 11.6
	\end{tabular}
	\caption{Experimentally determined tilt angle $\psi_{0}$, kink angle $\phi_{0}$, Area difference $\lvert\Delta A_{\mathrm{pz}}\rvert$ and stress intensity factors $\Delta K_{\mathrm{i},\psi,\mathrm{corr}}$ for plastic zone contours pictured in figure \ref{fig:Konturvergleich_PZ}}
	\label{tab:Num-exp Abgleich}
\end{table*}

Figure \ref{fig:Konturvergleich_PZ} shows an example of the experimentally and numerically determined plastic zones at $K_{\mathrm{I}}$ for both L-T (a,b) and T-L (c,d) orientation according to Table \ref{tab:Num-exp Abgleich}. For a better comparison, the coordinates $x_{pz}, y_{pz}$ are normalized towards the crack tip position. In terms of overall size and shape, there is good agreement between experimental and numerical results. At higher $K_{\mathrm{I}}$ the asymmetry and thus the wing area difference $\lvert\Delta A_{\mathrm{pz}}\rvert$ within the shape is more pronounced, as shown qualitatively in the contour in Figure \ref{fig:Konturvergleich_SIFS} d). A similar observation was made by Gao et al. that also found a deviation of the contour of the plastic zone from the shape of the dog bone in the strain field \cite{Gao.2022}. 
The contour detection algorithm described in subsection \ref{Plastic zone evaluation} has revealed separate contours. Figure \ref{fig:Konturvergleich_PZ} b) and d) show artefacts near the plastic zone with a significant size deviation between the detected contours. 
The equal size of the contours in Figure \ref{fig:Konturvergleich_PZ} a) and c) indicates crack bifurcations, which reduce the size of the plastic zone of the primary crack.
A similar conclusion was drawn by Bouiadjra et al. that analysed the influence of micro cracks on size and shape of the plastic zone. They stated that the occurrence of micro cracks caused by crack branching, changes the shape of the plastic zone and can lead to a bifurcation of its plastic zone's contour \cite{BachirBouiadjra.2008}. This leads to the conclusion that a high resolution of the plastic zone in the HR-DIC strain field can identify crack bifurcations. The comparison between experimental and numerical analyses also confirms the dependence of the asymmetry of the plastic zone to the crack front tilt angle $\psi_{0}$ as also outlined for the numerical model in subsection\ref{Parameterstudy} and visualized in Figure \ref{fig:Konturvergleich_SIFS}.

\subsection{Relationship between experimentally determined crack tilt angle $\psi_{0}$ and plastic zone shape difference $\Delta A_{\mathrm{pz}}$}

\begin{figure}[ht]
    \centering
    \includegraphics[width=7.8 cm]{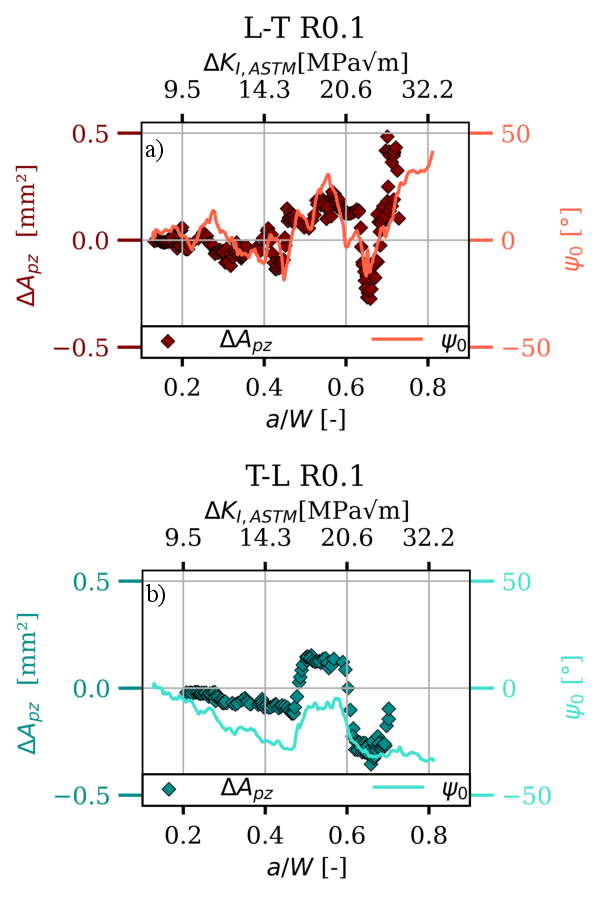}
    \caption{Comparison of tilt angle $\psi_{0}$ and plastic zone area difference $\Delta A_{pz}$ for L-T (a) and T-L specimen (b)}
    \label{fig:Zusammenhang_Verdrehwinkel_APZ}
\end{figure}

Figure \ref{fig:Zusammenhang_Verdrehwinkel_APZ} compares the tilt angle $\psi_{0}$ from fracture surface analysis with the wing area difference of the plastic zone $\Delta A_{pz}$ on the specimen surface from the HR-DIC system. Here, $\psi_{0}$ follows $\Delta A_{pz}$ perfectly and confirms the relationship between both descriptors that have been outlined in section \ref{tab:FE-Parameterstudie} and visualized in Figure \ref{fig:Flächendifferenz}.
We conclude, that surface-based, 2D information based on the crack tip strain field can represent 3D crack front geometries. In standardized fatigue crack growth experiments, the fracture modes and three-dimensional features are usually neglected, and the understanding of their effects towards crack propagation behaviour is limited \cite{Pook.2013}. By numerically identifying and experimentally confirming the relationship between three- and two-dimensional crack-describing quantities, a better understanding of three-dimensional effects regarding crack propagation behaviour can be obtained.

\section{Conclusions}
The HR-DIC data of the plastic zone provided time-series data of a propagating fatigue crack in AA2024-T3 sheet material. After the test, 3D scans of the fracture surface revealed its local topography. Various shape descriptors were used to derive a local numerical description of both the plastic zone and the surface topography. Based on this data, a new methodology was developed that relates 2D surface information to 3D thickness effects during fatigue crack growth, including the equivalent stress intensity factor. The following conclusions can be drawn:

\begin{itemize}
\item Temporal and high-resolution DIC data shows great potential to detect and analyze fatigue crack effects. The data can be correlated to the macroscopic crack behavior and compared with data from FE analyses to extend to 3D aspects of fatigue cracks. To distinguish between scatter and true effects, it is necessary to acquire HR-DIC data at least every 0.5mm of crack propagation.
\item Crack tilting leads to a mixed-mode condition of the crack front being represented by an asymmetric plastic zone. Therefore, we conclude that the plastic zone shape at the specimen surface represents 3D aspects of fatigue crack growth.
\item The mixed mode condition that occurs during crack tilting leads to deviation of the crack tip stress state with decrease of $K_{\mathrm{I}}$ and increase of $K_{\mathrm{III}}$, compared to a crack under pure mode I load. The stress intensity factors can be calculated by taking into account the twisting angle that is given by the fracture surface analysis or the asymmetry of the plastic zone.
\item Correlations can be made between the tilt angle and variations in the y-coordinates of the crack tip position, as well as the plastic zone ratio.
\item The rolling directions L-T and T-L influence the crack surface topology. By considering the local tilt angle in in the stress intensity factors, the calculation of an equivalent stress intensity factor results in almost superimposed crack propagation curves.
\end{itemize}

\bibliography{main}

\end{document}